\documentclass[]{article}

\usepackage{amsmath}
\usepackage{amsmath}
\usepackage{amssymb}
\usepackage{physics}
\usepackage{appendix}

\usepackage{blindtext} 
\usepackage{bookman}
\usepackage[T1]{fontenc} 
\usepackage{microtype} 

\usepackage[english]{babel} 

\usepackage[hmarginratio=1:1,top=32mm,columnsep=20pt]{geometry} 
\usepackage[hang, small,labelfont=bf,up,textfont=rm,up]{caption} 
\usepackage{booktabs} 

\usepackage{lettrine} 

\usepackage{enumitem} 
\setlist[itemize]{noitemsep} 

\usepackage{abstract} 

\usepackage{titlesec} 
\titleformat{\section}[block]{\large\scshape\centering}{\thesection.}{1em}{} 
\titleformat{\subsection}[block]{\large}{\thesubsection.}{1em}{} 

\usepackage{fancyhdr} 
\usepackage{titling} 
\usepackage{cite}
\usepackage{hyperref} 
\hypersetup{
    colorlinks=true,
    linkcolor=blue,
    filecolor=magenta,      
    urlcolor=blue,
    citecolor=blue,
    pdftitle={Overleaf Example},
    pdfpagemode=FullScreen,
    }

\usepackage[dvipsnames]{xcolor}
\usepackage[affil-it]{authblk}
\usepackage{graphicx}

\title{Emergent dark matter}
\author{%
\textsc{Christian Canete and Archil Kobakhidze} 
\vspace{0.2cm} \\
\normalsize \itshape
Sydney Consortium for Particle Physics and Cosmology, \\
\normalsize  \itshape
School of Physics, The University of Sydney, NSW 2006, Australia \\ 
\normalsize \itshape 
 \href{mailto:christian.canete@sydney.edu.au}{christian.canete@sydney.edu.au}, 
\href{mailto:archil.kobakhidze@sydney.edu.au}{archil.kobakhidze@sydney.edu.au}.

}
\date{} 
\renewcommand{\maketitlehookd}{%

\begin{abstract}
\noindent 
We entertain the possibility that the phenomena typically attributed to dark matter may have a fundamentally emergent nature, rather than arising from new particle degrees of freedom. To illustrate this idea, we consider a field-theoretic model of a three-form gauge field coupled to a cosmological fluid composed of ordinary matter and radiation. In the absence of interactions, the 3-form gauge theory describes only a global, non-propagating state, which can be associated with dark energy. However, when coupled to the cosmic fluid, the theory gives rise to an emergent, dynamical in-medium state. We identify this emergent state of the 3-form gauge field with dark matter. Thus, our proposal provides a unified framework for the dark sector of the universe within the context of an interacting three-form gauge theory. Furthermore, we speculate that the three-form field may have a gravitational origin, potentially supported by the lepton-number asymmetry in the primordial plasma. If this scenario is correct, conventional direct and indirect searches for dark matter would likely be futile.
\end{abstract}

}

\begin{document}

\maketitle


\section{Introduction}

The microscopic nature of dark matter remains unresolved. From a particle physics perspective, dark matter is thought to consist of particle states, similar to ordinary luminous matter. However, no known particles in the Standard Model have the properties required to form dark matter. As a result, solving the dark matter puzzle typically involves extending the Standard Model with new hypothetical particles and interactions. Non-particle candidates, like primordial black holes or topological and non-topological composite objects, also necessitate extending the Standard Model with new particle states. Likewise, alternative theories to the particle dark matter paradigm, such as modified gravity proposals, also rely on introducing new local propagating degrees of freedom.

In this paper, we propose a radically different perspective on the origin of dark matter. Specifically, we suggest that the effects attributed to dark matter are an emergent phenomena, not linked to any new fundamental particle degree of freedom. We explore this idea within a field-theoretic model based on the 3-form gauge theory.

The core of our proposal can be summarised as follows: a non-interacting 3-form gauge theory does not possess local propagating states (particles) in 4-dimensional space-time, as gauge redundancy removes all potential dynamical degrees of freedom. This holds even in the interacting theory provided gauge invariance is maintained and the 3-form gauge field remains massless. Importantly, though, it carries an energy that can be probed as a cosmological constant when coupled to gravity. This behaviour is dramatic when the 3-form gauge field develops an effective mass through the interaction with matter particles.  Namely, in a medium of ordinary matter, an analogue of the Anderson-Higgs phenomenon \cite{Anderson:1963pc} for the 3-form gauge field can occur. In the cosmological context, the energy stored in the 3-form gauge field is redistributed such that in-medium massive modes act as dark matter rather than dark energy.\footnote{Incidentally, the contribution of massive in-medium modes (plasmons) of the electromagnetic radiation to the energy density of the universe has been computed in \cite{plasmon}. It appears to represent a small fraction of the radiation energy density. In our framework, dark matter derives its energy from dark energy.} If this scenario is realised in nature, it would have profound implications for dark matter searches that critically rely on its particle nature — such experiments would be rendered futile.

We should note that the use of 3-form gauge theories have been employed in the literature to describe phenomena across a broad spectrum of physical systems, including confinement and the strong CP problem in QCD \cite{Luscher:1978rn, Aurilia:1980jz, Gabadadze:1997kj, Dvali:2005an, Dvali:2022fdv, Sakhelashvili:2021eid}, the emergence of mass hierarchies \cite{Dvali:2003br, Dvali:2004tma}, and the cosmological constant and inflation \cite{Hawking:1984hk, Brown:1987dd, Duff:1989ah, Bousso:2000xa, Dvali:2001sm, Kaloper:2008qs, Kaloper:2008fb}. Closer to our work, we mention the proposals in Refs. \cite{Ansoldi:2001xi} and \cite{Klinkhamer:2016zzh}. The first paper suggests that coupling a 3-form gauge field with fundamental 2-branes triggers the Anderson-Higgs mechanism. The second paper models a 3-form field with high derivative and potential terms, introducing a propagating degree of freedom from the outset. Both of these proposals differ conceptually and technically from our approach\footnote{A different framework for non-particle dark matter was proposed in \cite{Mukohyama:2009mz}, where the dark matter energy density emerges as an integration constant in a gravity theory without full diffeomorphism invariance \cite{Horava:2009uw} [for more recent similar work, see Ref. \cite{Kaplan:2023wyw}]. However, that theory also contains an extra propagating degree of freedom \cite{Kobakhidze:2009zr}. Similarly, models incorporating dark matter effects in extended diffeomorphism-invariant gravity theories (e.g., \cite{Deffayet:2024ciu}) also involve extra propagating degrees of freedom.}.

The rest of the paper is organised as follows. In the next section, we review the formulation of the 3-form gauge theory and introduce the field-theoretic description of a cosmic fluid.  In Section \ref{s3}, we couple 3-form and cosmic fluid via the gauge-invariant interactions.  Such a theory exhibits the Anderson-Higgs mechanism, where a longitudinal degree of freedom of the massive 3-form field emerges from the gauge-invariant interaction term, and the energy density of these interactions contributes to dark matter. In the concluding Section \ref{s4}, we summarise the key findings and ideas of this paper, along with a discussion on the possible origin of the 3-form gauge field.    

\section{Field-theoretic description of dark energy and the cosmic fluid}
\subsection{3-Form Gauge Field Description of Dark Energy}

Let us begin by recalling the key definitions and symmetries of a 3-form gauge theory. An Abelian 3-form gauge potential, $A_{\nu\rho\sigma}$, is a rank-3 antisymmetric tensor field that transforms under local gauge transformations parameterised by a rank-2 antisymmetric field $\Omega_{\nu\rho}$ in the folllowing way:

\begin{equation}
\delta_{\Omega} A_{\nu\rho\sigma} = \nabla_{\nu}\Omega_{\rho\sigma} + \nabla_{\rho}\Omega_{\sigma\nu} + \nabla_{\sigma}\Omega_{\nu\rho}~.
\label{transform}
\end{equation}

Not all components of $\Omega_{\nu\rho}$ are independent as the transformations (\ref{transform}) exhibit further gauge redundancies. Specifically, 2 out of the 6 components of the 2-rank tensor $\Omega_{\nu\rho}$ can be removed by a transformation parameterised by a vector field $\xi_{\nu}$ such that $\delta_{\xi}\Omega_{\nu\rho} = \nabla_{\nu}\xi_{\rho} - \nabla_{\rho}\xi_{\nu}$. Furthermore, $\xi_{\nu}$ itself is defined up to some harmonic scalar field i.e. $\xi_{\nu} = \nabla_{\nu}\theta$ where $\Box \theta = 0$. Consequently, only 4 independent gauge transformations remain which can gauge away all 4 components of the 3-form field, leaving no physical propagating degrees of freedom.


The first-order Lagrangian for a free 3-form gauge theory is given by\footnote{We prefer to work in the first-order formalism in this paper to avoid issues related to boundary conditions for field configurations that do not vanish rapidly at spatial infinity. For instance, properly accounting for boundary conditions when defining the on-shell energy-momentum tensor for 3-form gauge fields is discussed in \cite{Duff:1989ah}. In this formalism, both the 3-form gauge field $A_{\nu\rho\sigma}$ and its 4-form field strength $F_{\mu\nu\rho\sigma}$ are treated as independent fields. The second equation in (\ref{solform}), which defines the relation between 4-form field strength and 3-form gauge potential in the second-order formalism, here arises as the equation of motion obtained by varying the Lagrangian with respect to $F_{\mu\nu\rho\sigma}$. }  

\begin{equation}
\sqrt{-g}\mathcal{L}_{3f} = \frac{\sqrt{-g}}{2 \cdot 4!} F_{\mu\nu\rho\sigma} F^{\mu\nu\rho\sigma}-\frac{\sqrt{-g}}{3!}~F^{\mu\nu\rho\sigma}\nabla_{\left[\mu\right.}A_{\left.\nu\rho\sigma\right]}~,
\label{4formL}
\end{equation}
where $g$ is the determinant of the metric tensor and the square brackets denote taking the totally antisymmetric combination of tensors (see Appendix \ref{A} for definition). The equations of motion take the form
\begin{equation}
\partial_{\mu} \left(\sqrt{-g} F^{\mu\nu\rho\sigma} \right) = 0~, \quad F_{\mu\nu\rho\sigma}=4\nabla_{\left[\mu\right.}A_{\left.\nu\rho\sigma\right]} ~,
\label{solform}
\end{equation}
which admits a simple solution
\begin{equation}
F\equiv-\frac{1}{4!}\epsilon^{\mu\nu\rho\sigma}F_{\mu\nu\rho\sigma} = \sqrt{2}\Lambda ~,
\label{soll3form}
\end{equation}
where $\epsilon^{\mu\nu\rho\sigma}$ is the Levi-Civita tensor (see Appendix \ref{A} for our conventions) and $\Lambda$ is an arbitrary constant of dimension mass-squared. Therefore, the free 3-form gauge theory possesses only a single global degree of freedom - agreeing with our discussion on the gauge invariance of the theory. It is only this global degree of freedom that contributes to the energy-momentum tensor
\begin{align}
 T_{\mu\nu}^{3f}&= -\frac{1}{3!}F_{\mu\alpha\beta\gamma}F_\nu^{~\alpha\beta\gamma} +\frac{1}{2\cdot 4!}g_{\mu\nu}F_{\alpha\beta\gamma\delta}F^{\alpha\beta\gamma\delta} ~~\text{(off-shell)}~, 
 \label{em3form1}\\
 &=g_{\mu\nu}~\Lambda^2~~\text{(on-shell)}~.
 \label{em3form}
\end{align}
Hence, on-shell, the effect of the 3-form gauge field is indistinguishable from the effect of a cosmological constant term.

\subsection{A Field-Theoretic Model for Cosmic Fluid}

Let us now turn to the field-theoretic description of the cosmic fluid, which we assume to be an irrotational perfect fluid. Such a fluid can be described as a shift-invariant scalar field theory \cite{Madsen:1988ph} (see also \cite{Ballesteros:2016kdx} for a more general formulation) in a homogeneous cosmological spacetime. The relevant Lagrangian - in the first-order formalism - can be written as:

\begin{equation}
\sqrt{-g}\mathcal{L}_{cf} = \sqrt{-g}\left(-\mu^4 X^{(1+w_f)/2} + V^{\mu}\partial_{\mu}\phi\right)~,
\label{cf}
\end{equation}
where $X = \frac{1}{2\mu^4} g^{\mu\nu} V_{\mu} V_{\nu}$, and $g_{\mu\nu} = {\rm diag}[1, -a^2(t), -a^2(t), -a^2(t)]$ is the metric of a homogeneous and isotropic Friedmann–Lemaître–Robertson–Walker (FLRW) universe. Here, $\mu$ is a constant mass parameter defining the fluid’s energy density, and $w_f$ is a constant parameter specifying its equation of state.

Within the first-order formalism, $V_{\mu}$ is the one-form field strength of the scalar field $\phi$. Physically, the normalised one-form field corresponds to the fluid's 4-velocity
\begin{equation}
u^{\mu} = \frac{1}{\sqrt{2\mu^4 X}} V^{\mu}~.
\end{equation}
Since the fluid is irrotational, $V^{\mu}$ must be curl-free, i.e. $\epsilon^{\mu\nu\rho\sigma}\nabla_{\rho}V_{\sigma} = 0$. Furthermore, the invariance of the scalar field under global shift transformations ($\phi \to \phi + \text{constant}$) ensures the covariant conservation of the on-shell current

\begin{equation}
\nabla_{\mu} V^{\mu} \equiv \nabla_{\mu} \left(\sqrt{2\mu^4 X} u^{\mu}\right) = 0~.
\label{con}
\end{equation}

In FLRW spacetime, (\ref{con}) admits a simple solution that preserves the symmetries of the background
\begin{equation}
V^0 = \frac{\sqrt{2} c \mu^2}{a^3}~, \quad \vec{V} = 0~,
\label{s1}
\end{equation}
where $\sqrt{2} c \mu^2$ is an integration constant with $c$ dimensionless. Using this solution and referring to Eqs. (\ref{eqst}) and (\ref{eqst1}) in the Appendix \ref{B}, the energy density of the cosmic fluid is

\begin{align}
\rho_f &= \mu^4 X^{(1+w_f)/2} \nonumber \\
&= \frac{c^{1+w_f} \mu^4}{a^{3(1+w_f)}}~.
\end{align}
Here, the second line corresponds to the background energy density after substituting (\ref{s1}). The pressure is then given by $p_f = w_f \rho_f$.

The cosmic fluid considered here can be interpreted as part of the matter and/or radiation content of the universe, composed of ordinary Standard Model particles. For simplicity, we adopt a single-fluid approximation with a constant equation of state, $w_f = p_f / \rho_f$. Generalisation to a more complex interacting multi-fluid model is straightforward, but is not pursued here.

\section{Emergent dark matter in the 3-form gauge theory of cosmic fluid}
\label{s3}

We now arrive at a crucial step: coupling the cosmic fluid to a 3-form gauge theory. Our guiding principle is to preserve the symmetries of the free theories described above—namely, the local gauge invariance of the 3-form gauge theory and the global shift invariance of the cosmic fluid scalar field. This can be accomplished through the following interaction term alongside a Lagrange multiplier
\begin{equation}
\sqrt{-g}\mathcal{L}_{int}=\sqrt{-g}\frac{m}{3!} \epsilon^{\mu\nu\rho\sigma} V_{\mu}A_{\nu\rho\sigma} - \sqrt{-g}\frac{1}{3!}\epsilon^{\mu\nu\rho\sigma}\left(\nabla_{\mu}V_{\nu}\right)\Theta_{\rho\sigma}~,
\label{int}
\end{equation}
where $m$ is a coupling constant of dimension mass. The 2-form Lagrange multiplier $\Theta_{\rho\sigma}$ has the gauge transformation $\delta \Theta_{\rho\sigma} = \Omega_{\rho\sigma}$  to ensure the gauge invariance of the whole theory both off-shell and on-shell. Hence, our theory is described by the Lagrangian $\mathcal{L}=\mathcal{L}_{3f}+\mathcal{L}_{cf}+\mathcal{L}_{int}$.  

Note that the interaction term (\ref{int}) satisfies the desired symmetries and is topological in the sense that it has no dependence on the metric tensor. Such a topological term between $p$-form gauge field and the field strength of a $(d-p-1)$-form field in $d$-dimensions is known to be equivalent to the gauge-invariant theory of a massive $p$-form field. We refer the reader to Appendix \ref{C} for the explicit demonstration of this.\footnote{In 4-dimensions, this mechanism, known as the $BF$ topological mass generation, was first proposed for 1-form gauge fields in \cite{Cremmer:1973mg}, and for 3-form gauge fields in \cite{Dvali:2005an, Dvali:2005ws}. See discussion in \cite{new1} for general $p$-form gauge theory in $d$-dimensions.}

An immediate implication of the topological nature of the coupling between the cosmic fluid and 3-form gauge field for our cosmological model is that the \emph{off-shell} energy-momentum tensor does not involve the contribution from the interaction term (\ref{int}). Therefore, the off-shell energy-momentum tensor is a sum of the energy-momentum tensors for a free 3-form gauge theory (\ref{em3form1}) and that of the cosmic fluid (\ref{emfl}): $T_{\mu\nu}=T^{3f}_{\mu\nu}+T^{cf}_{\mu\nu}$. Nevertheless, the equations of motion are modified by the interaction term, and the on-shell energy-momentum tensor gets modified to
\begin{eqnarray}
T_{\mu\nu} &=& (1+w_f)\mu^4X^{(1+w_f)/2} u_{\mu}u_{\nu} - g_{\mu\nu}\left[w_f\mu^4X^{(1+w_f)/2} - \frac{1}{2}F^2 -\sqrt{2X}\mu^2mB^{\alpha}u_{\alpha}\right]
\nonumber \\
&-&\frac{1}{2}\sqrt{2X}\mu^2m\left(B_{\mu}u_{\nu} +B_{\nu}u_{\mu}\right)~,
    \label{em}
 \end{eqnarray}
 where we define a new field $B^{\mu}\equiv\frac{1}{3!}\epsilon^{\mu\nu\rho\sigma}\left(A_{\nu\rho\sigma} - \nabla_{\nu}\Theta_{\rho\sigma}\right)$ for convenience\footnote{In the absence of the Lagrange multiplier $\Theta_{\rho\sigma}$ this $B^{\mu}$ field is exactly the dual to the 3-form gauge field. The field strength tensor may also be expressed through this dual field $F=-\nabla_{\mu}B^{\mu}$.}. One observes that the last term in the above energy–momentum tensor represents a deviation from the perfect fluid description. Physically, this term describes the energy transfer between the two components of the model — the dark energy from the 3-form and the ordinary cosmic fluid. Consequently, the effective (time-dependent) equation of state that describes the combined 3-form and cosmic fluid takes the form:
\begin{eqnarray}
w_{eff}&=&\frac{w_f - \kappa}{1+\kappa}~, \quad\kappa =\frac{F^2}{2\mu^4X^{(1+w_f)/2}}~.
 \label{eqstate}
\end{eqnarray}
The parameter $\kappa$ in the above equation describes the relative energy densities carried by the 3-form gauge field and the cosmic fluid. If the energy stored in the 3-form field dominates over all other forms of energy, including the energy of the cosmic fluid ($\kappa\gg1$), then $\omega_{eff}\approx -1$ and one accounts for the dark energy-dominated universe. If the opposite is true ($\kappa\ll1$), then the universe can be dominated by the cosmic fluid with the equation of state $w_{eff}\approx w_f$. Thus, to describe the cosmology of the observable universe, one must start with a small $\kappa$ in the early universe (radiation/matter-dominated era) and transition to large values of $\kappa$ (present dark energy-dominated universe), i.e. $\dot{\kappa} > 0$ where the dot refers to a time derivative.   

To capture this dynamic, we inspect the background equations of motion for the combined system of a homogeneous 3-form gauge field and scalar field. We set $\Theta_{\rho\sigma} = 0$ and work in the Weyl-type gauge where $A_{0ij}=0$. The non-vanishing components of the 3-form gauge field can be parameterised through a real scalar field $\alpha(t)$
\begin{align}
A_{ijk}&=\alpha(t)\tilde \epsilon_{ijk}~, \\
\Longrightarrow F_{0123}=\dot{\alpha}&,~~F=-a^{-3}\dot\alpha~,
\end{align}
where $\tilde \epsilon_{ijk}$ is the 3-dimensional Levi-Civita numerical symbol. The equations of motion are then
\begin{align}
\label{eq1}
\partial_0\left(\frac{1}{a^3}\dot\alpha\right)&=m V_0~,  \\
\label{eq2}
\dot\phi - \frac{m}{a^{3}}\alpha &= \frac{\mu^2(1+w_f)}{\sqrt{2}}\left(\frac{V_0}{\sqrt{2}\mu^2}\right)^{w_f}~,
\end{align}
and are supplemented by the solution (\ref{s1}) for $V_0$. These equations, together with the Friedman equation, fully determine our cosmological model. 

The equation ({\ref{eq1}) above implies that the 3-form gauge field, in addition to the static component, admits a dynamical solution as well. Therefore, 3-form gauge theory no longer describes just the cosmological constant. Critical to this is the topological coupling (\ref{int}) that, as discussed, generates the massive mode of the 3-form gauge field. A general solution of the system of these equations is impossible to obtain in a closed form, while a numerical solution is not very illuminating. Therefore, we resort to a simplifying assumption. Namely, let us assume that at the early stages the equation of state is given by $w_{eff}\approx w_f$ ($\kappa \ll 1$), which then transitions to $w_{eff}\approx -1$ ($\kappa\gg1$), the regime of an accelerating universe.\footnote{As discussed in the literature \cite{Dvali:2013eja, Dvali:2014gua, Dvali:2017eba}, within the S-matrix formalism for quantum gravity, the eternal de Sitter universe is inconsistent. Therefore, there must be a mechanism that cancels the exact time-independent cosmological constant, while still allowing accelerated expansion of the universe (see, e.g., a scenario based on supergravity \cite{Dvali:2024dlb}).} In this approximation, we solved the dynamical equations and evaluated the on-shell energy-momentum tensor. In particular, the background energy density takes the following form
\begin{align}
    \bar{\rho} = \frac{\mu^4c^{1+w_f}}{a^{3(1+w_f)}} + \left\{
\begin{array}{ll}
      \left(\frac{1+w_{eff}}{1-w_{eff}}\frac{c\mu^2m}{H_0}a^{-3(1-w_{eff})/2} + \Lambda\right)^2 & \text{(cosmological fluid domination, $w_{eff}\approx w_f$}) \\
      \left(\frac{c\mu^2m}{3H_0}a^{-3} + \Lambda\right)^2 & \text{(3-form domination, $w_{eff}\approx -1$)}~,\\
\end{array} 
\label{ed}
\right.
\end{align}
where $H_0$ is the present value of the Hubble parameter. In our description, the 3-form dominates the present accelerated cosmological era. In this case, one can see that there are three components in the energy density in addition to the energy density described by pure cosmic fluid i.e the first term in (\ref{ed}). Let us first study the energy density components in the 3-form-dominated era. One of the new contributions describes that of a stiff fluid, i.e. the $c^2\mu^4m^2/9H_0^2$ term, which is diluting at a rate $\sim a^{-6}$. Next, we have a contribution that behaves as a non-relativistic fluid - this is what we identify as a potential dark matter candidate. If we assume that the total dark matter observed today originates directly from this 3-form interaction, then the the density goes like: $\rho_{dm,0}=2c\mu^2m\Lambda/3H_0$. Finally we have the dark energy density, $\Lambda^2$. Since the dark energy density must dominate in this era, the hierarchy of the above energy densities is straightforward, assuming $m\lesssim H_0$. At the same time, if we assume the cosmic fluid is dominated by baryonic matte i.e. $w_f\approx 0$, then we can associate the first term in the energy with the present-day baryonic energy density, $\rho_{b,0}=c\mu^4$. Recalling that $\rho_{dm,0}/\rho_{b,0}\approx 5$, we obtain the relation $m\approx \frac{15}{2}\frac{\mu^2H_0}{\Lambda}\approx \frac{15}{2\sqrt{3}}\frac{\mu^2}{M_P}$, where $M_P$ is the (reduced) Planck mass.

The earlier stages of the cosmic evolution may or may not be dominated by the one-component cosmic fluid. But if we assume it does dominate the cosmic expansion, the additional components behave as $a^{-3(1-w_f)}$, $a^{-3(1-w_f)/2}$ and the constant dark energy energy density, $\Lambda^2$, which is subdominant in the early universe. For $w_f\approx 0$, we can identify the dark matter energy density in this epoch as equal to $\rho_{dm, cf}=c^2\mu^4m^2/H_0^2$ alongside the fluid component that behaves as $a^{-3/2}$. The latter describes the effects of the energy transfer between the 3-form and cosmic fluid. Matching the dark matter densities at the matter-dark energy equality epoch again implies that the mass parameter describing the 3-form-fluid interactions is small i.e. $m\lesssim H_0$. Recall that this parameter defines the mass of the dynamical mode of the 3-form (see Appendix \ref{C}). For this reason, this dynamical mode cannot be resolved in traditional dark matter experiments.         

\section{Summary and discussion} 
\label{s4}

In this paper, we introduced the concept of emergent dark matter and discussed its realisation within a 3-form gauge theory coupled to the primordial plasma. A 3-form gauge theory coupled only to gravity is equivalent to introducing a non-dynamical cosmological constant, which can account for the dark energy content of the universe. When coupled in a gauge-invariant manner to the surrounding cosmological fluid—described using a shift-invariant scalar field—the 3-form gauge field acquires a dynamical mode. This mechanism is fully analogous to the emergence of massive plasmons in an electromagnetic plasma via the Anderson-Higgs mechanism. The energy of the 3-form gauge field is thus redistributed between the non-dynamical cosmological constant and the dynamical mode, which behaves as dark matter on cosmological scales. In the simplest model considered, the effective mass of this emergent mode is typically less than the present Hubble constant. Since there is no fundamental particle excitation associated with this dark matter, it cannot be detected through conventional dark matter detectors.

At the hydrodynamic level, small-scale behaviour, such as gravitational clumping, is straightforward to demonstrate. However, it remains an open question to better understand the small-scale structure of emergent dark matter, particularly given that it manifests only in the presence of ordinary matter and/or radiation.

On a more theoretical note, the origin of the 3-form gauge theory itself is intriguing. Higher-form gauge fields commonly appear in low-energy effective descriptions of string theory. In the spirit of this work, however, 3-form gauge fields can also be interpreted as effective descriptions of the vacua of gauge theories with topological structures, i.e., the so-called $\theta$-vacua. Such topological features appear in non-Abelian gauge theories governing the strong \cite{Luscher:1978rn, Aurilia:1980jz, Gabadadze:1997kj, Dvali:2005an} and weak \cite{Dvali:2024zpc, Dvali:2025pcx} interactions of the Standard Model, as well as in General Relativity \cite{Dvali:2024dlb}.

In the above scenario, the topological coupling (\ref{int})—which is central to our scenario and gives mass to the 3-form gauge field— may arise from a quantum anomaly associated with a classically conserved charge. Such a coupling may also emerge in a medium with a nonzero chemical potential for the anomalous charge. For example, in a primordial plasma with a net lepton number, a topological coupling to the gravitational Chern–Simons 3-form can be induced, with $V_{\mu}$ corresponding to the gradient of the chemical potential \cite{Barrie:2017mmr}. It would be worthwhile to further explore whether such mechanisms could provide a robust foundation for emergent dark matter.

\section*{Acknowledgments}
The authors gratefully acknowledge Gia Dvali and Oto Sakhelashvili for valuable discussions on various aspects of 3-form gauge theory. They also wish to thank the Max Planck Institute for Physics in Munich for hosting their visits, during which part of this work was conducted. This research was partially supported by the Australian Research Council through the Discovery Projects grant DP220101721. CC also acknowledges travel support from the University of Sydney’s PRSS and Grant-in-Aid schemes.
\appendix

\renewcommand{\theequation}{A-\arabic{equation}}
 \setcounter{equation}{0}

\section{Notations and convensions}
\label{A}
\renewcommand{\theequation}{A-\arabic{equation}}
 \setcounter{equation}{0}
 
We work with the mostly negative signature metric tensor. The 4-velocity $u_{\mu}$ is, therefore, normalised as $u^{\mu}u_{\mu}=1$, and the energy-momentum tensor is defined as
\begin{align}
    T_{\mu\nu} = \frac{2}{\sqrt{-g}}\frac{\partial\left(\sqrt{-g}\mathcal{L}\right)}{\partial g^{\mu\nu}} = 2\frac{\partial\mathcal{L}}{\partial g^{\mu\nu}} - g_{\mu\nu}\mathcal{L}~,
\end{align}
where $\mathcal{L}$ is the Lagrangian of the theory.

We work with tensors, rather than tensor densities, with the following convention for the Levi-Civita tensor
\begin{align}
    \epsilon_{0123} = \sqrt{-g}~, \quad \epsilon^{0123} = -\frac{1}{\sqrt{-g}}~.
\end{align}
For the reader's convenience, we provide some useful identities used in our calculations:
\begin{align}
\epsilon^{\mu\nu\rho\sigma}\epsilon_{\mu\beta\gamma\delta} &= -\delta^{\nu\rho\sigma}_{\beta\gamma\delta}~, \quad
    \epsilon^{\mu\nu\rho\sigma}\epsilon_{\mu\nu\gamma\delta} = -2\delta^{\rho\sigma}_{\gamma\delta}~, \\
    \epsilon^{\mu\nu\rho\sigma}\epsilon_{\mu\nu\rho\delta} &= -3!\delta^{\sigma}_{\delta}~, \quad 
    \epsilon^{\mu\nu\rho\sigma}\epsilon_{\mu\nu\rho\sigma} = -4!~.
\end{align}
where $\delta^{\mu_1...\mu_n}_{\alpha_1...\alpha_n}$ is the generalised Kronecker-delta, which acts on a tensor in the following way:
\begin{align}
\delta^{\mu_1...\mu_n}_{\alpha_1...\alpha_n}a^{\alpha_1...\alpha_n} = n!a^{[\mu_1...\mu_n]}~,
\end{align}
where $a$ is an arbitrary $n$-rank tensor and the square brackets $[\,\cdot\,]$ is shorthand notation to take the normalised totally antisymmetric combination of the tensor i.e.
\begin{align}
    a^{[\mu_1...\mu_n]} = \frac{1}{n!}\left(a^{\mu_1...\mu_n} + \{\text{all antisymmetric permutation of indices}\}\right)~.
\end{align}

\section{Cosmological fluid in the first-order formalism}
\label{B}
\renewcommand{\theequation}{B-\arabic{equation}}
 \setcounter{equation}{0}

As discussed in the main text, we describe the cosmological fluid by the shift-invariant scalar field ($\phi\to \phi+\text{constant}$) in the first-order formalism. The relevant Lagrangian to consider is:
\begin{align}
    \sqrt{-g}\mathcal{L}_{cf} = \sqrt{-g}\left[V^{\mu}\partial_{\mu}\phi - \mu^4 P(X)\right]~, \quad X \equiv \frac{1}{2\mu^4}g^{\mu\nu}V_{\mu}V_{\nu}~, \label{eq:cf_lagran}
\end{align}
with $\mu$ a parameter with dimensions of mass representing the energy scale of the theory. The equations of motion derived from varying $V_{\mu}$ and $\phi$, respectively, are
\begin{align}
      \partial^{\mu}\phi - P_XV^{\mu} = 0 ~, \quad  \label{eq:cf_eom_V}\partial_{\mu}\left(\sqrt{-g}V^{\mu}\right) = 0~,
\end{align}
where $P_X\equiv \delta P/\delta X$ is the functional derivative. The energy-momentum tensor of this theory is
\begin{align}
    T_{\mu\nu} = 2V_{\mu}\partial_{\nu}\phi - P_XV_{\mu}V_{\nu} - g_{\mu\nu}\left(V^{\alpha}\partial_{\alpha}\phi - \mu^4P\right)~~ \text{(off-shell)}~,
    \label{emfl}\\
    = 2\mu^4 XP_X u_{\mu}u_{\nu} - g_{\mu\nu}\left(2\mu^4 XP_X - \mu^4P\right)~~\text{(on-shell)}~,
\end{align}
where in the second line we have used the first of (\ref{eq:cf_eom_V}) and define the fluid 4-velocity as $u_{\mu}=V_{\mu}/\sqrt{2\mu^4X}$. The energy-momentum then takes the form of that of the perfect fluid with the equation of state
\begin{align}
    w_f = \frac{p_f}{\rho_f} = \frac{2XP_X}{P} - 1~.
    \label{eqst}
\end{align}
Solving this equation for $P(X)$, we obtain - up to an inconsequential constant factor
\begin{align}
    P(X) = X^{(1+w_f)/2} ~.
    \label{eqst1}
\end{align}
Hence, the action (\ref{cf}) employed in the main text follows. 

Note that this formalism allows us to describe a perfect fluid with an arbitrary equation of state $w_f$, including non-relativistic matter where $w_f=0$. If we had employed the second-order formalism, as is typically done in the literature, we would run into a problem. Indeed, the second-order formalism is obtained from (\ref{eq:cf_lagran}) by making the formal substitution $V_{\mu}=\partial_{\mu}\phi$. The Lagrangian density then becomes simply a functional of $X=\partial_{\mu}\phi\partial^{\mu}\phi/2\mu^4$ i.e. $\sqrt{-g}\mathcal{L}_{cf}=-\sqrt{-g}\mu^4 P(X)$. Consequently, the fluid pressure now reads $p_f=\mu^4 P(X)$. This framework obviously is not suitable for the description of the non-relativistic perfect fluid with $p_f=0$. For alternative descriptions, see Ref. \cite{Ballesteros:2016kdx}.

A similar inconsistency arises when using the second-order formalism for the 3-form gauge field without carefully accounting for boundary contributions \cite{Duff:1989ah}. The bulk Lagrangian in the second-order formalism reads as:  
$\mathcal{L}_{3f}=-\frac{1}{2\cdot 4!}F_{\mu\nu\rho\sigma}F^{\mu\nu\rho\sigma}.$ After substituting the solution (\ref{soll3form}), one obtains the on-shell Lagrangian, $\mathcal{L}_{3f}^{\text{on-shell}}=\Lambda^2$. The on-shell energy-momentum tensor obtained from this on-shell Lagrangian has a wrong sign, $T_{\mu\nu}=-g_{\mu\nu}~\Lambda^2$, c.f. (\ref{em3form}). This issue can be resolved if we introduce a boundary term of the form $\mathcal{L}_{bound}=(1/3!)\partial_{\mu}\left(\sqrt{-g}A_{\nu\rho\sigma}F^{\mu\nu\rho\sigma}\right)$ under the constraint that the field strength tensor vanishes at the boundary i.e. $\delta F_{\mu\nu\rho\sigma}|_{bound}=0$. Its on-shell value is $\mathcal{L}_{bound}^{\text{on-shell}}=-2\Lambda^2$, correcting the sign of the energy-momentum tensor. Alternatively, the first-order formalism constructed as in (\ref{4formL}) accounts for this discrepancy.

\section{Topological coupling and mass generation for 3-form gauge field}
\label{C}
\renewcommand{\theequation}{C-\arabic{equation}}
 \setcounter{equation}{0}
In this Appendix, we would like to demonstrate the generation of a massive propagating mode due to the topological coupling between a shift-invariant scalar field modelling the cosmic fluid and a 3-form gauge field. The total Lagrangian of the model reads: 
\begin{align}
    \mathcal{L}= -\frac{1}{2\cdot 4!}F_{\mu\nu\rho\sigma}F^{\mu\nu\rho\sigma} + V_{\mu}\partial^{\mu}\phi -\mu^4P(X) + \frac{m}{3!}\epsilon^{\mu\nu\rho\sigma}V_{\mu}A_{\nu\rho\sigma}-\frac{1}{3!}\epsilon^{\mu\nu\rho\sigma}\left(\nabla_{\mu}V_{\nu}\right)\Theta_{\rho\sigma}~,
    \label{full}
\end{align}
where the last term enforces the curl-free nature of the vector field $V_{\mu}$ through the Lagrange multiplier $\Theta_{\rho\sigma}$. Under this constraint, the above Lagrangian (and the equations of motion that follow) are explicitly invariant under the 3-form gauge transformations (\ref{transform}).

First, we note that the Hodge decomposition of the 3-form gauge field (ignoring the harmonic component) is 
\begin{align}
    A_{\nu\rho\sigma} = \frac{1}{m}\nabla_{\left[\nu\right.}\Omega_{\left.\rho\sigma\right]}-\frac{1}{m}\epsilon_{\mu\nu\rho\sigma}\partial^{\mu}\chi~.
\end{align}
The gauge-dependent, longitudinal part $\Omega_{\rho\sigma}$ can be removed from any gauge-invariant combination involving the 3-form - in particular the choice $\Omega_{\rho\sigma} = \Theta_{\rho\sigma}$ eliminates the Lagrange multiplier. Therefore, only the gauge-invariant, transverse part shows up explicitly through the scalar field $\chi$. In particular, the 4-form field strength $F=-\frac{1}{4!}\epsilon^{\mu\nu\rho\sigma}F_{\mu\nu\rho\sigma}$ takes the form $F=-\frac{1}{m}\Box \chi$. Consequently, the equation of motion for 3-form gauge field, 
\begin{equation}
\nabla_{\mu}F= - mV_{\mu}~,
    \label{eqn3f}
\end{equation}
along with $\nabla_{\mu}V^{\mu}=0$ implies that the field $\chi$ is a bi-harmonic scalar field: $\Box^2 \chi=0$.

On the other hand, the variation of the Lagrangian (\ref{full}) with respect to $V_{\mu}$ gives:
\begin{align}
\nabla^{\mu}(\phi+\chi)&=P_XV^{\mu}~, \nonumber \\
&=\frac{P_X}{m^2}\nabla^{\mu}\Box \chi~. \label{eom_V}
\end{align}
In principle, this equation can be solved for $V_{\mu}$ and then all the equations can be expressed in terms of scalar fields $\phi$ and $\chi$. However, the resulting equations, being highly non-linear and containing non-standard kinetic terms, are not particularly tractable. Instead, we expand $P(X)$ around a certain background solution $X=\bar{X}=\text{constant}$, and consider field fluctuations at time scales $\delta t < 1/H_0$ (flat spacetime approximation). We can thus expand $P_X$ around this background solution, which is best expressed in its Taylor series expansion using (\ref{eqst1})
\begin{align}
    P_X = \frac{1+w_f}{2}X^{(w_f-1)/2} = \frac{1+w_f}{2}\bar{X}^{(w_f-1)/2} + \frac{1+w_f}{2}\sum_{k=1}^{\infty}\left.\frac{\delta^kP_X}{\delta X^k}\right|_{X=\bar{X}}\left(X-\bar{X}\right)^k ~.
\end{align}
The second term of the RHS proportional to $(X-\bar{X})^k \sim \left(\nabla_{\mu}\Box\chi\nabla^{\mu}\Box\chi\right)^k$ contains higher-derivative operators that do not contribute to the mass term. Thus, we are left with
\begin{align}
    \frac{1+w_f}{2}\bar{X}^{(w_f-1)/2}\nabla^{\mu}\Box\chi = m^2\nabla^{\mu}\left(\chi + \phi\right) + \{\text{higher-derivative terms}\}~.
\end{align}
In the limit that we consider field oscillations much smaller than the Hubble expansion, we can approximate $a\sim1$ today, and so we have the background solution that $\bar{X} = c^2$. We thus have
\begin{align}
    \Box\chi = \frac{2c^{1-w_f}}{1+w_f}m^2\left(\chi + \phi\right) + \{\text{higher-derivative terms}\}~,
\end{align}
demonstrating that, indeed, we have a massive field $\chi$ with mass $m_{\chi}^2 = \frac{2c^{1-w_f}}{1+w_f}m^2$. For $\omega_f\approx 0$ considered in the main text, $m_{\chi}^2\sim \rho_B/M_P^2\ll H_0^2$. Hence, for any practical purpose, this excitation is massless at present epoch.  



\end{document}